\documentstyle[aps,prd,twocolumn]{revtex}

\let\ssection=\section
\renewcommand{\section}{\setcounter{equation}{0}\ssection}

\begin{document}
\draft

\title{}


%
\twocolumn[

%
]
\narrowtext
{\bf COMMENT ON:
``Integrable Systems in Stringy Gravity"}\\

This paper \cite{gal} is doubtless interesting, unfortunatly almost all
the material presented here is already known and has been published
elsewhere. Its first part is a standard introduction of generation
techniques in General Relativity given even in text books like
ref. \cite{kramer}. The main problem of this generation techniques
is to find an explicite expression for the $g\in G$ element of equation
(9) in Gal'tsov's paper. This expression is known for some cases like
Einstein-Maxwell theory (EM) \cite{neuge} (for the Einstein-Maxwell
plus Dilaton theory $g$ does not modify), n-dimensional gravity in
spacetime \cite{ma6}, n-dimensional gravity in potential space
\cite{maison} \cite{ma4}, etc. (see ref. \cite{ma26} for a review
of this topic). The first step for the solution of this problem in
potential space is to  find the form of the target space, see equation
(15) in Gal'tsov's paper.  But this metric is known for the
Einstein-Maxwell-Dilaton gravity  since 1969, see chapter I in ref.
\cite{habil},  where the metric is written just as in equation (15) of
Gal'tsov's paper, but using a parameter $2/(\xi-3/2)$ instead of
$\alpha^2$ of Gal'tsov's. After a Legendre transformation, this metric
has been intensively used for finding new exact solutions in
five-dimensional gravity (FDG) \cite{ma1} \cite{ma5} \cite{ma24}, where
$\alpha^2=3$ in Gal'tsov's notation, or $\xi=13/6$ in Neugebauer's
notation. Furthermore, this metric has been used for generating exact
static solutions in Dilaton gravity for $\alpha$ arbitrary in
\cite{ma36} (using the $SL(2,R)/U(1)$ representation of $g$), where we
apply the harmonic map ansatz \cite{ma15} \cite{ma29},  because we even
knew that the target space is non-symmetric for arbitrary  $\alpha$,
which implies that the Lax pair  is useless for the non-static case. In
the static case some identifications  between EM and FDG can be done, and
solutions for one of them can be mapped  into the other one, as mentioned
at  the end of paper \cite{ma24}. Therefore, the generalization for
arbitrary $\alpha$ was trivial for static fields. To see this, it is
sufficient to write down the matrix $g\in SL(3,R)$ of equation (4) of
ref. \cite{ma4} for magnetostatic fields, i.e. $\psi=\epsilon=0$
\begin{equation}
 g={1\over f\kappa^{2\over 3}}\left(\matrix{
-2f^2&            0        &           0          \cr
 0   &            2        &-{1\over\sqrt{2}}\chi \cr
 0   &-{1\over\sqrt{2}}\chi&-{1\over 4}(\chi^2-f\kappa^2)
}\right)
\end{equation}
or for electrostatic fields, i.e. $\chi=\epsilon=0$
\begin{equation}
 g=-\kappa^{4\over 3}\left(\matrix{
-2(\psi^2-{f\over \kappa^2})&           0         &-{1\over\sqrt{2}}\psi \cr
                 0        &  {2\over f\kappa^2} &           0           \cr
 -{1\over\sqrt{2}}\psi    &           0         &      -{1\over 4}
}\right).
\end{equation}
After a symmetry transformation, the $SL(2,R)/U(1)$ sectors of matrices
(1)  and (2) are just the matrix (20) in Gal'tsov's paper, with the
following identification given in eq.(7) of ref. \cite{ma36}
\[
\kappa^2=e^{-2\alpha\phi},
\]
and substituting $u=\chi$ for magnetostatic and $u=\psi$ for
electrostatic fields. Soliton solutions of the $SL(2,R)/U(1)$ sectors
are given in \cite{tesis} (see also ref. \cite{ma14}). Nevertheless, we
have found that the harmonic map method is more powerful  and convenient
for this kind of differential equations \cite{ma36}.

Most of the last part of Gal'tsov's paper has been published recently by
A. Garcia, O. V. Kechkin and Gal'tsov himself \cite{alberto}
\cite{kech}. Again, the target space (23)-(24) of Gal'tsov's  paper, and
the isometries analysis of the metric in the potential  space have been
carried out in ref. \cite{kech} section III systematically.

To end this comment I must say that although I have used  some references
of difficult access here, Mr. Gal'tsov cannot argue that  he did not know
this references. During his stay in Mexico last year (which  is
surprisingly not mentioned in his paper!!) I  personally lend him
reference \cite{habil}, gave him references  \cite{ma4}, \cite{ma5},
\cite{ma14}, \cite{ma24}, \cite{ma36},  \cite{ma26}, and he had easy
access to the other ones.

I thank Hernando Quevedo for helpful  discussions. This work is partially
supported by CONACYT-Mexico.\vspace{.2cm}

Tonatiuh Matos,
Departamento de F\'{\i}sica, Centro de  Investigaci\'on y de Estudios
Avanzados del I.P.N.,
Apdo. Post. 14-740, 07000, M\'exico, D.F., M\'exico.
tmatos@fis.cinvestav.mx.


\end{document}